\documentclass[aps,prl,10pt,twocolumn,showpacs,superscriptaddress]{revtex4-1}
\usepackage{graphicx}

\begin{document}

\title{
Hydrogen molecule spectrum by many-body $GW$ and Bethe-Salpeter equation
}

\author{Jing Li}
\affiliation{Universit\'e Grenoble Alpes, CEA, Leti, F-38000, Grenoble, France}
\author{Valerio Olevano}
\affiliation{Universit\'e Grenoble Alpes, F-38000 Grenoble, France}
\affiliation{CNRS, Institut N\'eel, F-38042 Grenoble, France}
\affiliation{European Theoretical Spectroscopy Facility (ETSF)}

\date{\today}

\begin{abstract}
We check the \textit{ab initio} $GW$ approximation and Bethe-Salpeter equation (BSE) many-body methodology against the exact solution benchmark of the hydrogen molecule H$_2$ ground state and excitation spectrum, and in comparison with the configuration interaction (CI) and time-dependent Hartree-Fock methods.
The comparison is made on all the states we could unambiguously identify from the excitonic wave functions' symmetry.
At the equilibrium distance $R = 1.4 \, a_0$, the $GW$+BSE energy levels are in good agreement with the exact results, with an accuracy of 0.1$\sim$0.2~eV.
$GW$+BSE potential-energy curves are also in good agreement with the CI and the exact result up to $2.3 \, a_0$.
The solution no longer exists beyond $3.0 \, a_0$ for triplets ($4.3 \, a_0$ for singlets) due to instability of the ground state.
We tried to improve the $GW$ reference ground state by a renormalized random-phase approximation (r-RPA), but this did not solve the problem.
\end{abstract}


\maketitle

\textit{Introduction.---}Hydrogen H$_2$ is the simplest neutral molecule and one of the most straightforward many-body systems in nature.
In contrast to the hydrogen atom, where the exact analytical solution is known, H$_2$ already faces the quantum many-body problem to calculate electronic correlations.
Due to the presence of two electrons and their many-body interaction, the closed-form solution of the Schr\"odinger equation for H$_2$ does not exist.
Nevertheless, thanks to James and Coolidge's pioneer work \cite{ JamesCoolidge33}, the methodology devised originally by Hylleraas \cite{Hylleraas29} for the helium atom, which provides an exact solution in a numerical analysis sense, was adapted to H$_2$.
By exploiting the rotational symmetry around the dimer axis, the H$_2$ wave function can be written as a power series of five coordinates (instead of the three helium Hylleraas coordinates), that is the elliptic confocal coordinates $\xi_1, \xi_2, \eta_1, \eta_2$ and the electrons distance $\rho$.
In analogy with the hydrogen and the helium atom, for H$_2$ an exponential on the elliptic radial coordinates $\xi_1$ and $\xi_2$ is also introduced to speed up the series convergence.
The solution is then searched by varying the series coefficients up to a given order. Meanwhile, the next order can be used to evaluate the absolute error.
By increasing the order of the series, the error can be arbitrarily reduced.
Today, the H$_2$ solution is known up to an accuracy of $10^{-15}$ \cite{Pachucki10,PachuckiKomasa16}.
Beyond providing a rigorous way to validate theory against more and more accurate experiments \cite{HolschMerkt19,PuchalskiPachucki19}, this exact numerical result makes H$_2$ an ideal workbench to check any approximate many-body methodology.

In this work, we use the H$_2$ exact solution benchmark to check the \textit{ab initio} many-body methodology of the $GW$ approximation on the self-energy and the resolution of the Bethe-Salpeter equation (BSE) \cite{MartinReiningCeperley}.
For this purpose, a comparison is done against the exact solution of the idealized nonrelativistic and clamped nuclei H$_2$ Hamiltonian, excluding nuclear motion, relativistic and QED radiative corrections, and other complications not related to the many-body problem to calculate correlation energies.
We also compare with other more or less accurate many-body approaches, from full configuration interaction (CI) \cite{HelgakerJorgensenOlsen} down to Hartree-Fock (HF), passing through time-dependent HF [TDHF, also known in nuclear physics as random-phase approximation (RPA) \cite{RingSchuck} or RPA with exchange diagrams (RPAx)], and finally an approach known as renormalized RPA (r-RPA) \cite{CataraVanGiai96,CataraSambataro98,SchuckTohyama16,LiOlevano19,SchuckTohyama20}.
The comparison is made on all the states that it was possible to identify.
Our results show that at the H$_2$ equilibrium distance of $R = 1.4$ $a_0$, the $GW$+BSE energy levels are in good agreement with the exact results, with an accuracy of 0.1$\sim$0.2~eV, which is four times better than TDHF.
The $GW$+BSE energy-potential curves as a function of the nuclei distance are also remarkable up to at least $R = 2.3$ $a_0$, after which they start to deteriorate.
Beyond $R = 3.0$ $a_0$, we have no more $GW$+BSE solutions due to ground state instability.
This problem is not solved by r-RPA.

\textit{Methods.---} The starting point of our \textit{ab initio} many-body calculation is a standard HF calculation.
One can also start from density functional theory (DFT), e.g., in the local-density approximation or something else, but we opted for the zero-correlation, more meaningful, physical HF for our comparisons.
This is also the most standard for isolated systems.
We used a \emph{d-aug}-cc-pV5Z \cite{Dunning94} correlation-consistent Gaussian basis set with angular momentum up to $l=5$ and a double set of diffuse orbitals (105 Gaussians per atom and a total of 210 basis elements for the molecule) for all our calculations.
On top of HF, we performed a $GW$ self-energy contour-deformation calculation with self-consistency only on quasiparticle energies using a Coulomb-fitting resolution of the identity (RI-V) with the associated auxiliary basis \emph{d-aug}-cc-pV5Z-RI \cite{Wei06}.
The last step was a BSE calculation beyond the Tamm-Dancoff approximation of the excitation energies $\Omega_\lambda$ and the excitonic wave functions $\Psi_\lambda$ by diagonalization of the excitonic Hamiltonian,
\begin{equation}
 \left(
 \begin{array}{cc}
   A & B \\
   -B^* & -A^* \\
 \end{array}
 \right)
 \Psi_\lambda = \Omega_\lambda \Psi_\lambda
 , \label{bseeq}
\end{equation}
with
\begin{eqnarray}
  A &=& (\epsilon_p - \epsilon_h) \delta_{pp'} \delta_{hh'} + w_{ph'hp'} - W_{ph'p'h} 
  , \label{A} \\
  B &=& + w_{pp'hh'} - W_{pp'h'h}
  \label{B} ,
\end{eqnarray}
where $\epsilon_h$ ($\epsilon_p$) are hole (particle) [occupied (empty)] $GW$ quasiparticle energies, and $w$ ($W$) are bare (screened) Coulomb interaction matrix elements between $GW$ states $\phi_i$, e.g., $w_{ijkl} = \langle \phi_i(r) \phi_j(r') | w(r,r') | \phi_k(r) \phi_l(r') \rangle$.
The ground-state energy was calculated by the formula \cite{traceformula}
\newcommand{\tr}{\mathop{\mathrm{Tr}}}
\begin{equation}
 E_0 = E_0^\mathrm{HF} + \frac{1}{2} \left( \sum_\lambda \Omega_\lambda - \tr(A) \right)
 . \label{trace}
\end{equation}
We used the codes \texttt{NWCHEM} \cite{nwchem} and \texttt{ORCA} \cite{orca} for the HF and CI calculations, and  \texttt{Fiesta} \cite{BlaseOlevano11,Jac15a,Li16} with some checks by 
\texttt{TurboMole} \cite{turbomole} for $GW$, BSE, r-RPA, and TDHF.
\begin{figure}[t]
  \includegraphics[width=\columnwidth]{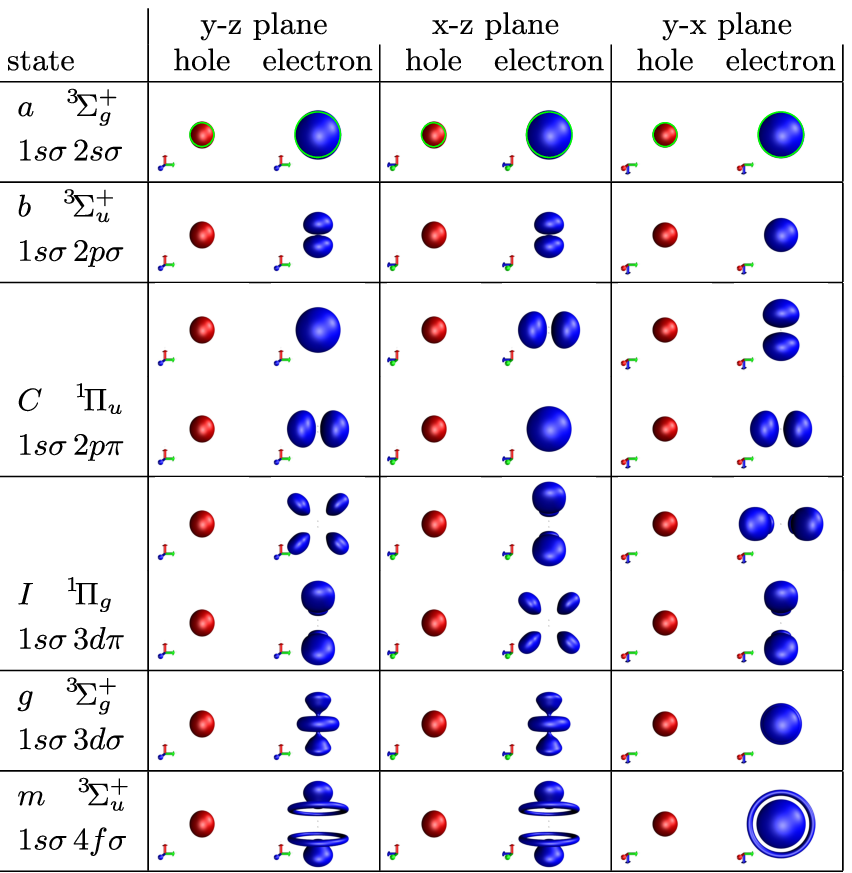}
\caption{
   Plot of selected excitons of H$_2$ obtained by BSE: the triplets $a$, $b$ (which is unbound), $g$ and $m$, and the singlet $C$ and $I$.
   On the left and in red ($1s\sigma$-like shapes): hole probability density distribution, $\rho_\lambda(r_h) = \int dr_e \Psi_\lambda^2(r_h,r_e)$; on the right and in blue: electron probability density distribution, $\rho_\lambda(r_e) = \int dr_h \Psi_\lambda^2(r_h,r_e)$.
   The two H atoms are indicated as small dots and their axis is oriented along $z$.
   Perfect circles (in green) are drawn for the $a$ exciton on top of both distributions to appreciate their elongation along $z$.
 }
 \label{excitons}
\end{figure}

\textit{Results.---} In Fig.~\ref{excitons}, we plot the excitonic wave functions $\Psi_\lambda$ of the most intuitive excitons.
At the left and in red, we plot the hole probability density distribution, $\rho_\lambda(r_h) = \int dr_e \Psi_\lambda^2(r_h,r_e)$, and at the right and in blue the electron distribution,  $\rho_\lambda(r_e) = \int dr_h \Psi_\lambda^2(r_h,r_e)$, in the planes $xz$, $yz$, and $xy$.
Note that we are plotting the square of the wave function, so that the two colors, red and blue, refer to electron and hole probability densities, not to the phase $+/-$ information, which is lost.
The hole distribution is, in all cases, trivially the $1s\sigma$-like orbital of the ground state where the hole is dug.
It looks like a perfectly spherical 1$s$ atomic state, but it is elongated along the H-H molecular axis in reality, and similarly for electron distributions.
Perfect green circles traced on the $a$ exciton electron and hole clouds help emphasize this elongation.
The study of the symmetry of the excitonic wave functions is essential to the unambiguous identification of the H$_2$ excitations, including the less intuitive ones, to compare with the literature.

\begin{figure}[t]
 \includegraphics[width=\columnwidth]{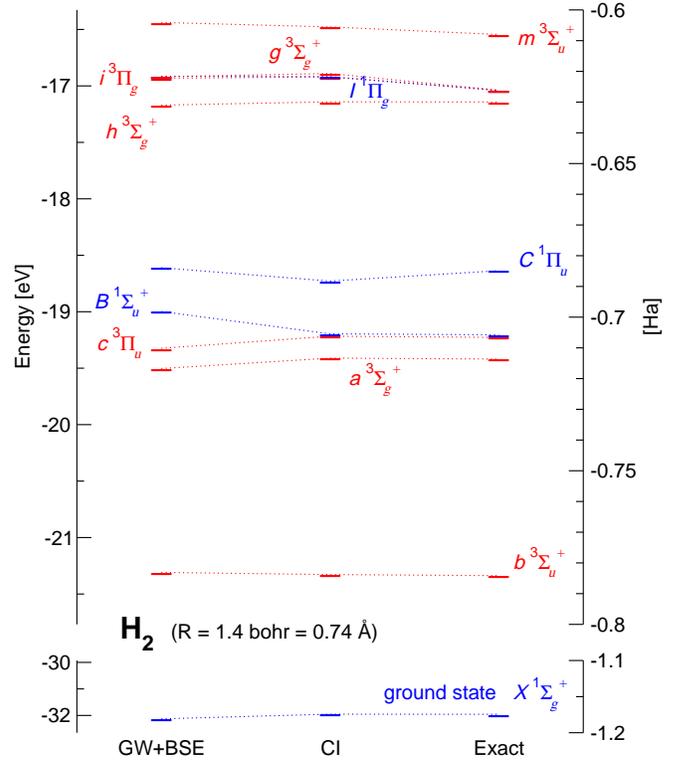}
 \caption{
   H$_2$ energy levels; plot of the results of Table~\ref{h2states}.
 }
 \label{h2plot}
\end{figure}

In Fig.~\ref{h2plot} and in Table~\ref{h2states}, we report the H$_2$ ground- and excited-state energy levels.
To identify the states, we use the notation by Dieke \cite{Dieke}, and we also indicate the united atoms (He) notation used by Sharp \cite{Sharp71}.
We report the exact levels from the literature \cite{KolosWolniewicz65,WakefieldDavidson65,KolosWolniewicz68,KolosWolniewicz68a,KolosWRychlewski77} which, as should be noted, are the solution to the clamped nuclei (at the equilibrium distance $R = 1.4 \, a_0$) H$_2$ nonrelativistic Schr\"odinger equation.
High-accuracy comparisons with the experiment should also consider nuclear motion, relativistic, and QED corrections
\cite{HolschMerkt19,PuchalskiPachucki19}.
We then report our TDHF, $GW$+RPAx, $GW$+BSE, and CI results, all calculated at the same \emph{d-aug}-cc-pV5Z basis level.
Full CI has a remarkable accuracy of $2 \cdot 10^{-4}$~Ha, at least up to the $h$ state, beyond which the performances of the \emph{d-aug}-cc-pV5Z basis start to deteriorate.
The error raises one order of magnitude ($5 \cdot 10^{-3}$~Ha).
The $GW$+BSE absolute error can be quantified at the level of 0.1$\sim$0.2~eV, thus confirming previous estimates for the $GW$ approximation.
Remarkably, the $GW$+BSE error is four times better than the TDHF (\textit{alias} HF+RPAx or also the RPA of nuclear physics).
The $GW$+RPAx shows an intermediate accuracy.
The route, TDHF $\to$ $GW$+RPAx $\to$ $GW$+BSE, demonstrates the improvement in accuracy step by step with the introduction of the correlations (on top of the uncorrelated HF) as accounted by screening, i.e., the screened Coulomb interaction $W$, in both the $GW$ self-energy, $\Sigma^{GW} = i G W$, and in the BSE kernel, $\Xi^\mathrm{BSE} = w - W$, of Eqs.~(\ref{A}) and (\ref{B}).

\begin{table}[t]
  \begin{tabular}{l @{} llccccl}
  \hline\hline
  \multicolumn{3}{l}{State}  & TDHF & \parbox[c][5ex]{.04\textwidth}{$GW$\\RPAx} & \parbox[c][5ex]{.04\textwidth}{$GW$\\BSE} & CI & \hfill \textbf{Exact} [Ha] \hfill\\ 
  \hline
  $X$ & $^1\!\Sigma^+_g$ & $1s\sigma$	& $-1.19$ & $-1.184$	& $-1.181$	& $-1.1743$	& $\mathbf{-1.17447571}$  \\ 
  $b$ & $^3\!\Sigma^+_u$ & $2p\sigma$	& $-0.83$ & $-0.818$	& $-0.783$	& $-0.7842$	& $\mathbf{-0.7841501}$  \\
  $a$ & $^3\!\Sigma^+_g$ & $2s\sigma$	& $-0.75$ & $-0.726$	& $-0.716$	& $-0.7135$	& $\mathbf{-0.7136358}$  \\
  $c$ & $^3\!\Pi_u$      & $2p\pi$	& $-0.73$ & $-0.716$	& $-0.708$	& $-0.7064$	& $\mathbf{-0.70658282}$ \\
  $B$ & $^1\!\Sigma^+_u$ & $2p\sigma$	& $-0.72$ & $-0.702$	& $-0.698$	& $-0.7056$	& $\mathbf{-0.7057434}$  \\
  $C$ & $^1\!\Pi_u$      & $2p\pi$	& $-0.70$ & $-0.685$	& $-0.683$	& $-0.6885$	& \\
  $C$ & $^1\!\Pi_u$      & $2p\pi$	& $-0.70$ & $-0.681$	& $-0.679$	& $-0.6848$	& $\mathbf{-0.6848598}$\footnote{
    The exact \cite{KolosWolniewicz65} result for the $C$ state refers to $R = 1.375$~$a_0$.}  \\
  $h$ & $^3\!\Sigma^+_g$ & $3s\sigma$	& $-0.65$ & $-0.633$	& $-0.631$	& $-0.6301$	& $\mathbf{-0.62995}$    \\
  $i$ & $^3\!\Pi_g$      & $3d\pi$	& $-0.64$ & $-0.622$	& $-0.622$	& $-0.6221$	& $\mathbf{-0.62623079}$ \\
  $I$ & $^1\!\Pi_g$      & $3d\pi$	& $-0.64$ & $-0.622$	& $-0.622$	& $-0.6220$	& $\mathbf{-0.62617190}$ \\
  $g$ & $^3\!\Sigma^+_g$ & $3d\sigma$	& $-0.64$ & $-0.622$	& $-0.622$	& $-0.6209$	& $\mathbf{-0.62611}$    \\
  $m$ & $^3\!\Sigma^+_u$ & $4f\sigma$	& $-0.63$ & $-0.607$	& $-0.604$	& $-0.6038$	& $\mathbf{-0.608}$      \\
  \hline\hline
  \end{tabular}
  \caption{H$_2$ energy levels [Ha] at the equilibrium distance $R = 1.4$~$a_0$.
    We compare the TDHF (\textit{alias} HF+RPAx), $GW$+RPAx, $GW$+BSE, and CI results, all calculated at the \emph{d-aug}-cc-pV5Z basis, to the exact results in the literature: $X$ \cite{Pachucki10}, $b$, $C$ \cite{KolosWolniewicz65}, $h$, $g$ \cite{WakefieldDavidson65}, $a$, $B$ \cite{KolosWolniewicz68}, $c$, $i$, $I$ \cite{KolosWRychlewski77}, $m$ \cite{Davidsonpriv}.
  }
  \label{h2states}
\end{table}

\begin{table}[b]
 \caption{
 H$_2$ ionization potential (eV).
 We report the HOMO energies for HF, single-iteration $G_0W_0$, and eigenvalue self-consistent ev$GW$.
 The exact value is the difference between the H$_2^+$  \cite{OlivaresPilonTurbiner18} and the H$_2$ \cite{KolosWolniewicz65} ground-state exact total energies calculated at $R = 1.4$ bohr (i.e., the vertical difference).  
 }
 \label{IP}
 \begin{tabular*}{\columnwidth}{@{\extracolsep{\fill}} lcccc}
  \hline\hline
  ($R = 1.4$ bohr) & HF & $G_0W_0$ & ev$GW$ & \textbf{Exact}\\
  \hline
  IP [eV] & 16.19 & 16.60 & 16.62 & \textbf{16.45} \\
  \hline\hline
 \end{tabular*}
\end{table}

In Table~\ref{IP}, we report another observable which has the exact value: the ionization potential (IP).
IP is the minimum energy to remove an electron from the H$_2$ molecule in its ground state, resulting in a H$_2^+$ ion plus an electron at infinity.
The H$_2^+$ ion is a one-electron system for which an analytical solution exists to the Schr\"odinger equation \cite{BatesReid68}.
The H$_2$ IP is defined as the difference between its ground-state energy $E_{X}^{\mathrm{H}_2} = -1.1745$ Ha and the energy $E_{X}^{\mathrm{H}_2^+}(R=1.4) = -0.5699$ Ha  \cite{OlivaresPilonTurbiner18} of the ground state of the H$_2^+$ ion \emph{at the H$_2$ equilibrium distance $R = 1.4$ bohr} (experimental measures do not leave the time to the nuclei to relax to the ion equilibrium distance), with the reversed sign.
This value is the exact IP, as reported in Table~\ref{IP}, which provides another essential check specific to the $GW$ approximation.
In HF and $GW$, the IP is the highest occupied molecular orbital (HOMO) energy with reversed sign.
One can see in Table~\ref{IP} that $GW$ has an error of 0.15$\sim$0.17 eV and improves the HF error of 0.26 eV.

\begin{figure}
 \includegraphics[width=\columnwidth]{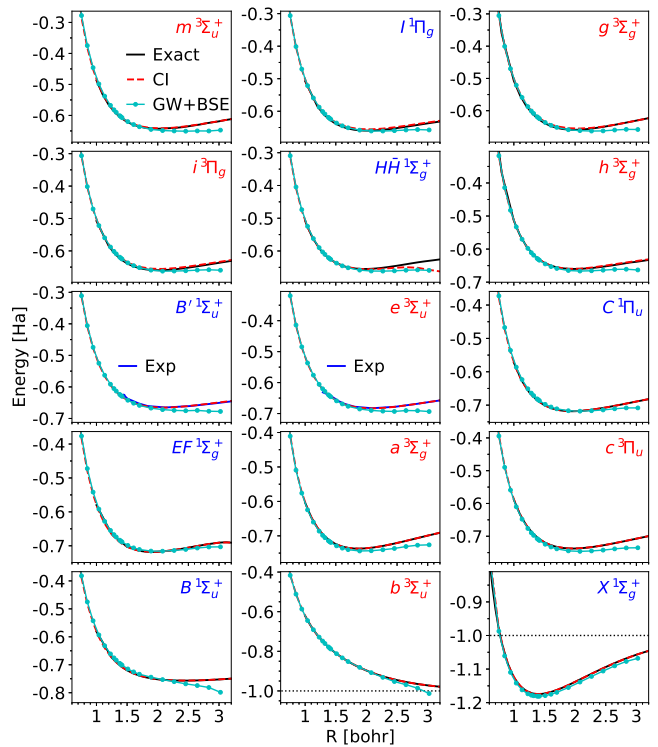}
 \caption{
   H$_2$ energy levels as a function of the H-H bond length: comparison between $GW$+BSE (cyan line with dots), CI (red dashes), and exact results in the literature (black line) for $m$ \cite{Davidsonpriv}, $I$, $i$, $c$  \cite{KolosWRychlewski77}, $g$, $h$ \cite{WakefieldDavidson65}, $H\bar{H}$ \cite{WolniewiczDressler85,WolniewiczDressler94}, $C$, $b$, $X$, \cite{KolosWolniewicz65,KolosWolniewicz68a}, $EF$ \cite{KolosWolniewicz69} $a$, $B$ \cite{KolosWolniewicz68}, or experiments (blue line) for $B'$ \cite{Spindler69} and $e$ \cite{Dieke58}.
 }
 \label{disscurv}
\end{figure}

In Fig.~\ref{disscurv}, we plot the energy as a function of the internuclear distance for all the states of Table~\ref{h2states} plus others for which we could find further data in the literature.
We again compare the $GW$+BSE curve to the CI and the exact result (the experiment for the $B'$ and $e$ states for which we could not find exact calculations in the literature).
The agreement with both CI and the exact result can be considered very good, at least up to $R = 2.3 \, a_0$, which is more than sufficient to capture the relevant range of the molecule binding.
Then it starts to deteriorate and, after $R = 3.0 \, a_0$, the $GW$+BSE solution no longer exists.
From this point on, a \textit{triplet instability} occurs: the diagonalization of the Bethe-Salpeter excitonic matrix provides imaginary eigenvalues, signaling that the reference ground state, i.e., the $GW$ one calculated (not fully self-consistently with respect to the wave functions) on top of the HF, is unstable toward another lesser energy ground state.
We can see in Fig.~\ref{disscurv} that the unbound triplet $b$~$^3\Sigma_u^+$ state, at the last point beyond 3~$a_0$, takes the value of $-1.01$~Ha, implying that it has already unphysically crossed the asymptote of $-1.0 \, \textrm{Ha} = -2.0$~Ryd, which is the exact analytic energy of two dissociated H atoms.
After that distance, the $GW$+BSE $b$~$^3\Sigma_u^+$ state tends to swap with the true ground state $X$~$^1\Sigma_g^+$ and becomes a spurious ground state with total spin $S = 1$.
For singlets, the instability occurs a bit further away, at $R = 4.3 \, a_0$, so that the agreement with the CI and exact results lasts a bit longer.
This would be even more evident if we compare excitation energy-differences with respect to the ground state, i.e., the $\Omega_\lambda$ directly out of Eq.~(\ref{bseeq}), instead of the absolute excitation energies of Fig.~\ref{disscurv}, i.e., the $E_\lambda = \Omega_\lambda + E_0$, which require evaluation of the ground-state energy $E_0$ by Eq.~(\ref{trace}) including triplets that are instable at shorter distance.
Note that the $^1\Sigma^+_g$ excited states, e.g., the $EF$  2s$\sigma$+2p$\sigma^2$, presents a secondary minimum at 4.39 $a_0$ and a relative maximum at 3.12 $a_0$ \cite{Sharp71,KolosWolniewicz69}.
It could be interesting to check the $GW$+BSE approach on this peculiar feature.
Unfortunately, the triplet instability occurs just immediately before.
The $GW$+BSE first derivative is approaching zero before the instability, and the excitation energy $\Omega_\lambda$ of Eq.~(\ref{bseeq}) runs on top of the CI until the instability, thus pointing to the relative maximum at least.
However, we cannot be conclusive on this point since the addition of the ground-state energy $E_0$, $E_\lambda = \Omega_\lambda + E_0$, plays an important role to conjure the double minimum, and the perceptible deviation toward lower energies of the $GW$+BSE ground state $E_0$ for $R>2.8 \, a_0$ (see $X$ $^1 \Sigma_G^+$ in Fig.~\ref{disscurv}) can contribute to nullify the first derivative of the $EF$ state.
The same holds for the $H\bar{H}$ $^1\Sigma^+_g$ 3s$\sigma$, but here the $\bar{H}$ secondary minimum occurs at a much larger distance, $R=11.2 \, a_0$ \cite{WolniewiczDressler85,WolniewiczDressler94}, and one can see that even CI is in trouble on this state due to degeneration with the closer $GK$ state of the same symmetry, $^1\Sigma^+_g$.

\begin{figure}[t]
 \includegraphics[width=\columnwidth]{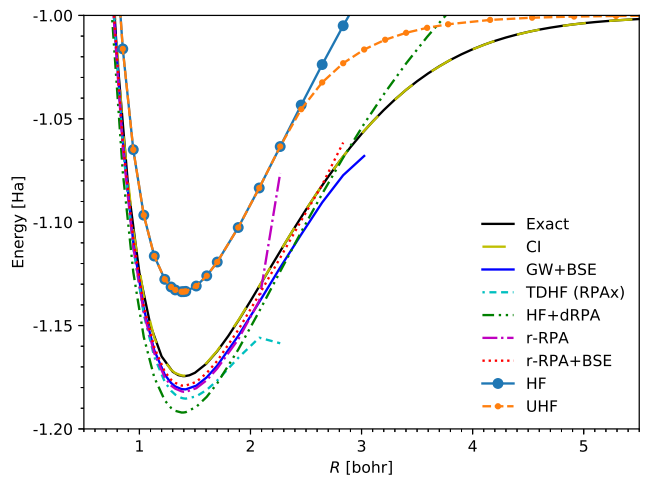}
 \caption{
   $X$~$^1\Sigma_g^+$ energy as a function of the H$_2$ bond length.
 }
 \label{groundstate}
\end{figure}

To solve the instability problem, one should look for a better reference ground state to start with, i.e.\ single quasiparticle energies $\epsilon_i$ and orbitals $\phi_i$ better than the $GW$ or HF.
A possibility can be the renormalized RPA (r-RPA) approach described in Refs.~\cite{CataraVanGiai96,CataraSambataro98,SchuckTohyama16,LiOlevano19,SchuckTohyama20}.
Compared to non fully self-consistent $GW$, which updates only quasiparticle energies, r-RPA updates both energies and occupation numbers. 
Such effect introduces some correlation on top of HF wave functions, so to have a better ground state and in the hope of pushing the instability to larger distances.
In principle, r-RPA should be carried up to self-consistency, typically three or four iterations.
Here we only did a single iteration to see whether there is already an improvement on TDHF (RPAx) and the HF reference ground state.
In Fig.~\ref{groundstate}, we report the results only for the ground state.
We first plot the HF uncorrelated result and the unrestricted HF (UHF), which coincides with restricted spin $S = 0$ HF up to the Coulson-Fisher point at 2.3~$a_0$.
We then report the $GW$+BSE result, which at the equilibrium distance is at 0.1$\sim$0.2 eV from the exact and the CI results.
The $GW$+BSE solution no longer exists beyond 3~$a_0$ (triplet instability), with problems already starting at 2.8~$a_0$.
For the TDHF approach, problems already start at 2.1~$a_0$, with no longer solution (triplet instability) after 2.3~$a_0$. 
We can say that the problem is more severe in TDHF (RPAx) than in $GW$+BSE, though for $GW$+BSE, a singlet instability appears after 4.3~$a_0$, whereas there is no singlet instability at any distance for TDHF.
On the other hand, there is no instability problem at all for the (direct) RPA calculation on top of HF (HF+dRPA): there is no singlet instability in dRPA, like in RPAx, while triplets energies keep at the uncorrelated level of HF energy differences in dRPA, so that they do not contribute to the ground-state correlation energy.
The HF+dRPA improves the HF curve, but the dissociation limit (not shown; see, e.g., Ref.~[\onlinecite{CarusoScheffler13}]) is still too large.
Finally, we report our r-RPA result which locates above TDHF (RPAx), with an improvement which almost achieves the same accuracy of $GW$+BSE.
But also for r-RPA, the solution does not exist beyond 2.3~$a_0$: on this point, r-RPA does not improve on TDHF (RPAx).
Still we can consider the $GW$+BSE reference ground state better, even though in this approach we only update quasiparticle energies and not the wave functions also, like in r-RPA.
To clarify this point, we performed a hybrid r-RPA+BSE calculation consisting in the use of the r-RPA approach to update both energies and also occupation numbers and wave functions, together with the use of the BSE kernel, $\Xi^\mathrm{BSE} = w - W$, with the screened Coulomb interaction $W$ [see Eqs.~(\ref{A}) and (\ref{B})], instead of the bare Coulomb $w$ of the TDHF kernel $\Xi^\mathrm{TDHF} = w - w$ [obtained replacing $W$ with $w$ in Eqs.~(\ref{A}) and (\ref{B})] and of the simplified r-RPA approach, which does not update the kernel shape.
The result, as shown in Fig.~\ref{groundstate}, is not that bad.
r-RPA+BSE improves the agreement with the exact result at the equilibrium distance, and the triplet instability occurs at almost the same distance as of the $GW$+BSE approach.
In principle, in fully self-consistent RPA (SCRPA) \cite{SchuckTohyama20} calculations beyond simplified r-RPA, the kernel is also updated and should start to contain screening.
Nevertheless, the hybrid r-RPA+BSE is not very well justified from an analytic perspective.
We also observe that the hybrid curve manifests a strange crossing with the exact results curve, which looks quite unphysical.

\textit{Comparison with previous work.---} To the best of our search in the literature, we could only find the results of Ref.~[\onlinecite{ReboliniSavin13}] as relevant for our study of H$_2$ excitations. 
Their results refer to a minimal basis set and cannot be directly compared to the real experimental and exact H$_2$.
Nevertheless, our and their results are coherent qualitatively (see their Fig.~1 with nonexact $G_0$).
On the H$_2$ ground state, the literature is vaster.
Our HF+dRPA curve practically coincides with the RPA@HF curve of Ref.~[\onlinecite{CarusoScheffler13}], though we used the trace formula (TF) Eq.~(\ref{trace}) and they used the adiabatic-connection fluctuation-dissipation theorem (ACFDT) $\lambda$-integration.
This confirms that the two formulas are equivalent in the direct RPA case, as demonstrated \cite{Furche08}.
In the other cases, i.e., TDHF (RPAx) or $GW$+BSE, the two formulas are not equivalent and their results may differ.
Thus the TDHF curve by ACFDT, called HF-RPA in Ref.~[\onlinecite{HesselmannGorling11prl}], differs from our TDHF by TF
\footnote{Rigorously \cite{HesselmannGorling11}, the ACFDT formula is exact only in references where the density is equal to the exact density all along the adiabatic connection $\lambda$ path, $\rho_\lambda = \rho_\mathrm{exact} \, \forall \lambda$, that is, only in Kohn-Sham DFT/TDDFT and for jellium in the homogeneous electron gas phase.}. 
Triplets are included in the TF Eq.~(\ref{trace}), whereas they do not contribute to ACFDT TDHF, resulting in larger energy.
On the other hand, the advantage is the absence of the triplet instability problem; in TDHF ACFDT, the solution exists up to dissociation.
ACFDT should be in trouble for $GW$+BSE where the instability also occurs for singlets. 
In Ref.~[\onlinecite{OlsenThygesen14}], they took into account only the $R$ and the $\lambda$ where a real solution exists for singlets, and neglected all imaginary poles.
Although our $GW$+BSE correlation energy by TF is more accurate at the equilibrium distance, their approximation was successful in describing the dissociation limit.
This strategy evidently cannot help here, where we are interested in excited states.

\textit{Conclusions.---} On the benchmark of the H$_2$ exact result, the $GW$+BSE many-body approach achieves, at the equilibrium distance, an accuracy of 0.1$\sim$0.2 eV, which is four times better than the TDHF (RPAx) 0.5 eV error, on all the states up to $m$ $^3\!\Sigma^+_u$ 4f$\sigma$.
The $GW$+BSE energy-potential curves are in good agreement with CI and exact results at least up to 2.3~$a_0$ and stop at the triplet instability at 3~$a_0$.
Improvement of the $GW$ reference ground state by the r-RPA approximation, which updates not only energies but also occupation numbers and wave functions, does not improve on the triplet instability problem.
The introduction of screening in r-RPA, as by an r-RPA+BSE hybrid, improves the triplet instability, which occurs at a similar distance compared to  $GW$+BSE.

\textit{Acknowledgments.---} We thank X. Blase, I. Duchemin, and P. Schuck for useful discussions.

\textit{Note added.---} Recently, we were made aware of a recent work \cite{BergerRomaniello20} also presenting an H$_2$ ground state $GW$+BSE dissociation curve which, \textit{mutatis mutandis}, i.e., use of a different basis set, of single-iteration $G_0W_0$, and in particular of the ACFDT formula, is nevertheless more in agreement with our result than with Ref.~\cite{OlsenThygesen14}.
It also presents interesting results using the Coulomb hole plus screened exchange (COHSEX) approximation, which is a static approximation on top of $GW$.

\bibliography{h2}

\end{document}